# APE Results of Hadron Masses in Full QCD Simulations. *

S. Antonelli[a], A. Bartoloni[a], C. Battista[a], M. Bellacci[a], S. Cabasino[a], N. Cabibbo[a],
L.A. Fernández[b], E. Panizzi[a], P.S. Paolucci[a], A. Muñoz-Sudupe[b], J.J. Ruiz-Lorenzo[a],
R. Sarno[a], A. Tarancón[c], G.M. Todesco[a], M. Torelli[a], P. Vicini[a].

[a]INFN Sezione di Roma I and Dipartimento di Fisica, Universitá di Roma "La Sapienza", P. A.Moro, 00185 Roma, Italy,

[b]Departamento de Física Teórica, Universidad Complutense de Madrid, Ciudad Universitaria, 28040 Madrid, Spain,

[c]Departamento de Física Teórica, Universidad de Zaragoza, Pza. de San Francisco s/n, 50009 Zaragoza, Spain,

We present numerical results obtained in full QCD with 2 flavors of Wilson fermions. We discuss the relation between the phase of Polyakov loops and the **sea** quarks boundary conditions. We report preliminary results about the HMC autocorrelation of the hadronic masses, on a $16^3 \times 32$ lattice volume, at $\beta = 5.55$ with $k_{sea} = 0.1570$.

## 1. THE SPATIAL POLYAKOV LOOPS PHASE.

Polyakov loops are responsible for the difference between quenched and unquenched finite size effects on the QCD mass spectrum[1][2]. The reason can be understood by looking at the **valence** quark hopping parameter expansion of the meson propagator [2]:

$$G = \sum_C k_{val}^{l(C)} \langle W(C) \rangle + \sum_C k_{val}^{l(C)} \sigma_{val} \langle P(C) \rangle . \quad (1)$$

The sums extend over all closed paths ($C$) of length $l(C)$. $W(C)$ are standard Wilson loops, completely contained into the lattice, while $P(C)$ are Polyakov type loops and $\langle \cdot \rangle$ denotes gauge field average; the value of the index $\sigma_{val}$ depends on the spatial boundary conditions on the **valence** quarks: $\sigma_{val} = +1$ for periodic boundary conditions, PBC, and $\sigma_{val} = (-1)^n$ for antiperiodic boundary conditions, APBC, respectively, with $n$ the number of windings around the lattice.

In full QCD the fermionic part of the action (both Wilson and staggered fermions) breaks explicitly the simmetry that, in quenched confined QCD guarantees $<P> = 0$. A simple model[3] may be useful to clarify the situation. The pres-

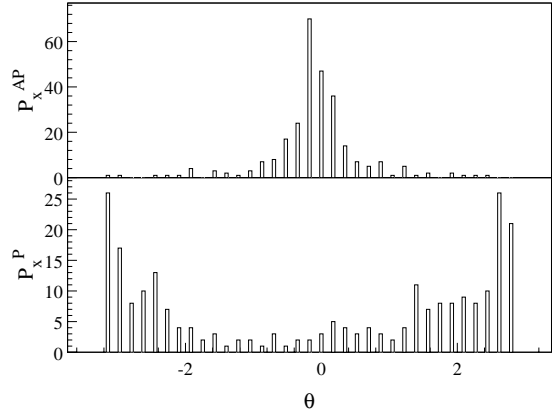

Figure 1. Histogram of the phase, $\theta$, of the $x$ component of the Polyakov loop, $P_x^{\text{AP}}$, for APBC and $P_x^{\text{P}}$, for PBC on **sea** quarks.

---
*Talk presented by R.Sarno

1

ence of the fermionic part of the QCD action is analogous to the existence of a magnetic field $h$ which breaks the $Z_3$ symmetry in a model for a spin, $\Pi$, which can take the three possible $Z_3$ values. $\Pi$ is coupled to the external magnetic field via the Hamiltonian

$$H_h = h\Pi + h^\dagger \Pi^\dagger . \qquad (2)$$

$H_h$ is not $Z_3$ invariant. The values taken by $\Pi$ are in correspondence with the expected phases of Polyakov loops, while the values of $h$ are in correspondence with the chosen **sea** quarks boundary conditions. Following this model we expect that with PBC on **sea** quarks the $e^{i2\pi/3}$ and the $e^{-i2\pi/3}$ phases of the Polyakov loops will be present with equal probability, while Polyakov loops are likely to be polarized in the $\Pi_0$ direction if APBC are used. Similarly we expect to be able to align the Polyakov loops along the $e^{i2\pi/3}$ (or $e^{-i2\pi/3}$) in the $Z_3$ space, if we choose $-e^{-i2\pi/3}$ (or $-e^{i2\pi/3}$ respectively) boundary conditions on the **sea** quarks.

We checked this suggestions with a simulation, performed on APE100, with 2 flavor Wilson fermions at $\beta = 5.3$ on a $8^3 \times 32$ lattice with $k_{sea} = 0.1670$. We carried out two different runs, one with fully PBC on the **sea** quarks and the other one with APBC in the spatial directions and PBC in the temporal one. Gauge configurations have been produced with APE100 using the HMC algorithm[4]. We created a set of 440 (thermalization) + 1350 thermalized trajectories. We have performed the measurement of the spatial loops, $P_x, P_y$ and $P_z$, on one every 5 consecutive trajectories, using the smearing procedure[5] to reduce fluctuations. With APBC on spatial directions, the phases are close to zero, while with PBC, the phases are concentrated in two regions, near $e^{i2\pi/3}$ and $e^{-i2\pi/3}$, as we can see from fig. 1.

Fixing the Polyakov loop phase reduces the statistical fluctuations on the hadron propagator and, hence, on the computed hadron masses. Furthermore the use of PBC on sea quarks causes a *flip-flop* effect between the two phases of the Polyakov loops, $e^{-i2\pi/3}$, $e^{i2\pi/3}$, which can produce a longer thermalization time.

## 2. SIMULATIONS ON $16^3 \times 32$ AT $\beta = 5.55$.

We started with a run on a $16^3 \times 32$ lattice volume at $\beta = 5.55$ with $k_{sea} = 0.1570$. We use APBC on spatial directions on **sea** quarks. We created a set of 900 (thermalization) + 500 trajectories. The last 500 have $dt = 0.083$ and $N_{MD} = 60$ ($dt \times N_{MD} \sim 0.5$). We have performed a measure, every 10 trajectories, of the hadronic correlation functions, of the $0^{++}$ and $2^{++}$ glueball correlation functions and of the Polyakov loop. For each trajectory we estimate the average value of the Wilson loop, $W_{1\times 1}$, (the plaquette). In the calculation of the valence quark propagator we use a smeared source of size $7^3$, while for the glueball correlation functions we use 21 iterations of the smearing. The physical lattice size (estimated from ref.[6]) is $L \sim 1.5$ fm.

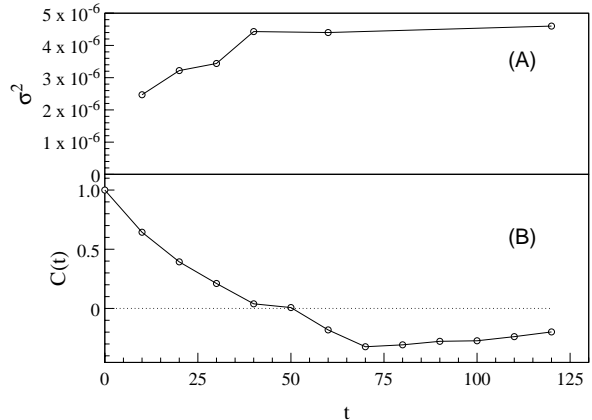

Figure 2. Autocorrelation for the pseudoscalar, $\pi$, meson correlation functions at the Euclidean time $t = 16$. On the top we present the results from method (A) and on the bottom the ones from method (B).

The first goal that we want to reach is an understanding of the correlations between measures done on consecutive HMC trajectories. This study is important in order to estimate a correct value for the statistical error. We use two dif-ferent methods: A) we study the behavior of the jacknife error $\sigma^2$ (calculated by dividing the measures in N bins of size T) as a function of the size T, analyzing when the error reaches a plateau and, B), we calculate the autocorrelation function $C(t)$ at time $\tau_s$ (see eq.(2.19) of ref.[7]).

Table 1
Values of the hadronic masses, measured with $k_{valence} = k_{sea} = 0.1570$, using APBC on spatial directions on the **sea** quarks, PBC on the **valence** quarks and a $7^3$ smeared source. We report also the value for the square ratio of the pseudoscalar to the vector mesons and the value for the plaquette.

| $\pi$ ($\gamma_5$) | $\rho$ ($\gamma_i$) | Nucleon | $N^*$ |
|---|---|---|---|
| 0.479(5) | 0.567(7) | 0.89(1) | 1.32(5) |
| $\Delta$ | $\Delta^\star$ | $(\frac{m_\pi}{m_\rho})^2$ | $W_{1 \times 1}$ |
| 0.95(1) | 1.36(8) | 0.71(1) | 0.5635(1) |

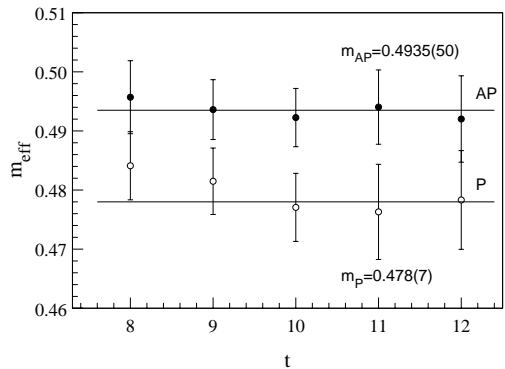

Figure 3. Pseudoscalar effective mass with both PBC and APBC on **valence** quarks, from propagators without smeared source.

With both methods we obtain for the plaquette and the hadrons correlation functions, at Euclidean time 16 (the middle of the temporal lattice), an autocorrelation time of about 50 HMC trajectories (see fig. 2). Starting from these result we estimate our statistical errors with the jacknife method used on 10 bin of 5 measures (50 trajectories) each. Results are reported in Table 1. We can see from this table that the masses are estimated with a statistical error of about 1%.

In order to study the systematic errors coming from the finite lattice, we calculated the hadronic masses with both PBC and APBC on **valence** quark. The difference between the values of masses obtained gives us an estimation of the contribution from $<P> \neq 0$ of eq. 1. From fig. 3, we can see an error of 3% on the estimation of the plateau of the $\pi$ and of 6% for the $\rho$. The systematic finite size errors are greater than the statistical errors quoted in Table 1. This difference is lower than that found for staggered fermions in fig 2 of ref.[2] as expected since our lattice is almost a factor 2 larger. However the effect is in the same direction: if the boundary conditions for **sea** quarks are AP, the masses obtained with PBC on **valence** quarks are lower than those obtained with APBC on **valence** quarks, as ex-pected from eq. (1). This is true for both Wilson and staggered fermions.

We want to stress that the smearing procedure is crucial also in full QCD, as we have observed in the study of the baryons correlation functions.

Acknowledgements

We thank G. Parisi for many suggestions and useful discussions. We thank for support and discussions E. Marinari, F. Marzano, F. Rapuano, R. Tripiccione and F. Zuliani.